\documentstyle[twocolumn,prb,aps,epsf]{revtex}

\begin{document}
\draft
\title{Off-equilibrium dynamics in a singular diffusion model}
\author{Federico Corberi$^1$, Mario Nicodemi$^2$, Marina Piccioni$^1$ 
and Antonio Coniglio$^1$}
\address{
$^1$ Dipartimento di Scienze Fisiche, Universit\`a di Napoli and
Istituto Nazionale di Fisica della Materia, Unit\`a di Napoli,
Mostra d'Oltremare, Pad. 19, 80125 Napoli, Italy  \\
$^2$ Department of Mathematics, Imperial College,
Huxley Building, 180 Queen's Gate, London, SW7 2BZ, U.K.}

\maketitle

\begin{abstract}

We introduce a schematic non-linear diffusion model where density fluctuations
induce a rich out of equilibrium dynamics.
The properties of the
model are studied by numerical simulations
and analytically in a mean field approximation. 
At low temperatures and high densities we find 
a long off-equilibrium glassy region, where the system evolves out 
of an initially pinned state showing aging and a slow decay 
of the autocorrelation as an enhanced power law, along with strong 
spatial heterogeneities and violation of the fluctuation dissipation theorem. 
\end{abstract}


As fluids are supercooled below the melting temperature
their structural relaxation becomes very slow and may result
in a glass transition characterized by a dramatic increase of 
the viscosity. 
Under these conditions the dynamical evolution of glass-forming
liquids shows a markedly out of equilibrium behavior \cite{Angell}.
In the main approach to the glassy dynamics, the Mode Coupling 
theory \cite {gotze}, 
the dynamical equations are solved
by resumming a non trivial set of diagrams.
This theory predicts an equilibrium relaxation time $\tau _0$ which
diverges as a power law at a dynamical transition.
Today it's well
established \cite{Angell} that the quoted theory applies in a region located
well before the point of structural arrest,
where the relaxation time is found experimentally to diverge according to the 
Voghel-Fulcher law \cite{Angell}
\begin{equation}
\tau _0 \sim \exp [v(\rho _c - \overline \rho )^{-1}]
\label{tau0}
\end{equation} 
Therefore we do not have information from
the theoretical point of view on the kinetics   
close to the dynamical transition where standard theories do not apply. 
In this paper we introduce a schematic diffusion
equation with a phenomenologically chosen mobility
which reproduces the equilibrium relaxation time (\ref{tau0})
observed in this region.
We then study the consequences of such an 
assumption on the out of equilibrium dynamics.
Due to its relative simplicity the model is amenable of both
numerical and analytical investigations allowing a complete
description of its features. Despite the simplicity of the equilibrium 
properties, the out of equilibrium
behavior is complex and in qualitative agreement with the known 
properties of real systems. Our analytical results
could in principle be tested quantitatively in experimental data and in
molecular dynamics simulations.

Glassy systems are usually schematized as composed of particles rattling 
inside ``cages" of typical size $a$ formed by the neighbors. 
Although diffusion is unimpeded inside the cells, motion
over larger distances is strongly suppressed at 
high molecular densities because a global rearrangement of many particles
is required. This glassy behavior is known to be reproduced also in
the simple case of hard spheres \cite{Angell}.
For suitably coarse graining of space and time scales,
the dynamics of supercooled liquids is Brownian in its microscopic
origin.
Therefore we consider a diffusion equation
for the variable $\rho (\vec r,t)$ which represents the coarse grained
particle density over distances of order $a$

\begin{equation}
\frac {\partial \rho (\vec r,t)}{\partial t}=\nabla \left [
M(\rho)\nabla \frac{\delta F \{\rho \}}{\delta \rho}\right ]
+\eta (\vec r,t)
\label{diff}
\end{equation}

In Eq.~(\ref{diff}) $F[\rho]=\int d\vec r \left[
\rho \ln \rho +(1-\rho)\ln (1-\rho )\right]$ 
is the entropy of the lattice gas-like model \cite{FH}, which we consider 
for sake of simplicity. 
More realistic forms of $F[\rho ]$, which take into account the attractive
interactions between particles can also be considered; here we find,
however, that the qualitative features of the model do not change \cite{lungo}.
$M(\rho )$ is a mobility, specified below, 
which is supposed to capture the main features of the constrained cooperative
dynamics of a dense fluid. 
$\eta (\vec r,t)$ is a gaussianly distributed random field,
representing the thermal noise, 
whose expectations are given \cite{corr} by $<\eta (\vec r,t)>=0$, 
$<\eta (\vec r,t)\eta(\vec r',t')>=-2T\nabla \left \{
M(\rho )\nabla \left [ \delta (\vec r-\vec r')\delta (t-t')  
\right ] \right \}$, where $<\cdots>$ means ensemble averages and
$T$ is the temperature in units of the Boltzmann constant $k_B$.

Now we specify $M(\rho)$. As shown below, from Eq.~(\ref{diff}) 
the characteristic equilibrium relaxation time $\tau _0$ behaves
as $M ^{-1}(\overline \rho )$, $\overline \rho $ being the average density,
for low $T$.
Then, in order to reproduce the behavior (\ref{tau0}) of $\tau _0$,
we assume a local
mobility of the form \cite{Ben}
\begin{equation}
M(\rho)=e^{v[\rho (\vec r,t) -1]^{-1}}
\label{mob2}
\end{equation} 
where a rescaled density, so that $\rho _c =1$, has been considered.
Due to its generality, 
Eq.~(\ref{diff}) is also suited for the description
of different physical systems where a constrained cooperative dynamics
is believed to play a fundamental role, such as granular materials 
\cite{NCprl1}.
We have studied Eq.~(\ref{diff}) for a temperature quench, by simulations
and in a mean field approach. We sketch our numerical results before 
entering a detailed mean field analysis.

Eq.~(\ref{diff}) has been simulated by a standard first order Euler 
discretization scheme on a $128$x$128$ two-dimensional square lattice with 
periodic boundary conditions, starting from an uncorrelated high temperature
initial state. The system is quenched to a very low temperature (results 
will be
presented for $T=10^{-4}$, but similar behaviors are found for different
temperatures).

\begin{figure}
\centerline{\epsfxsize=.9\hsize \epsffile{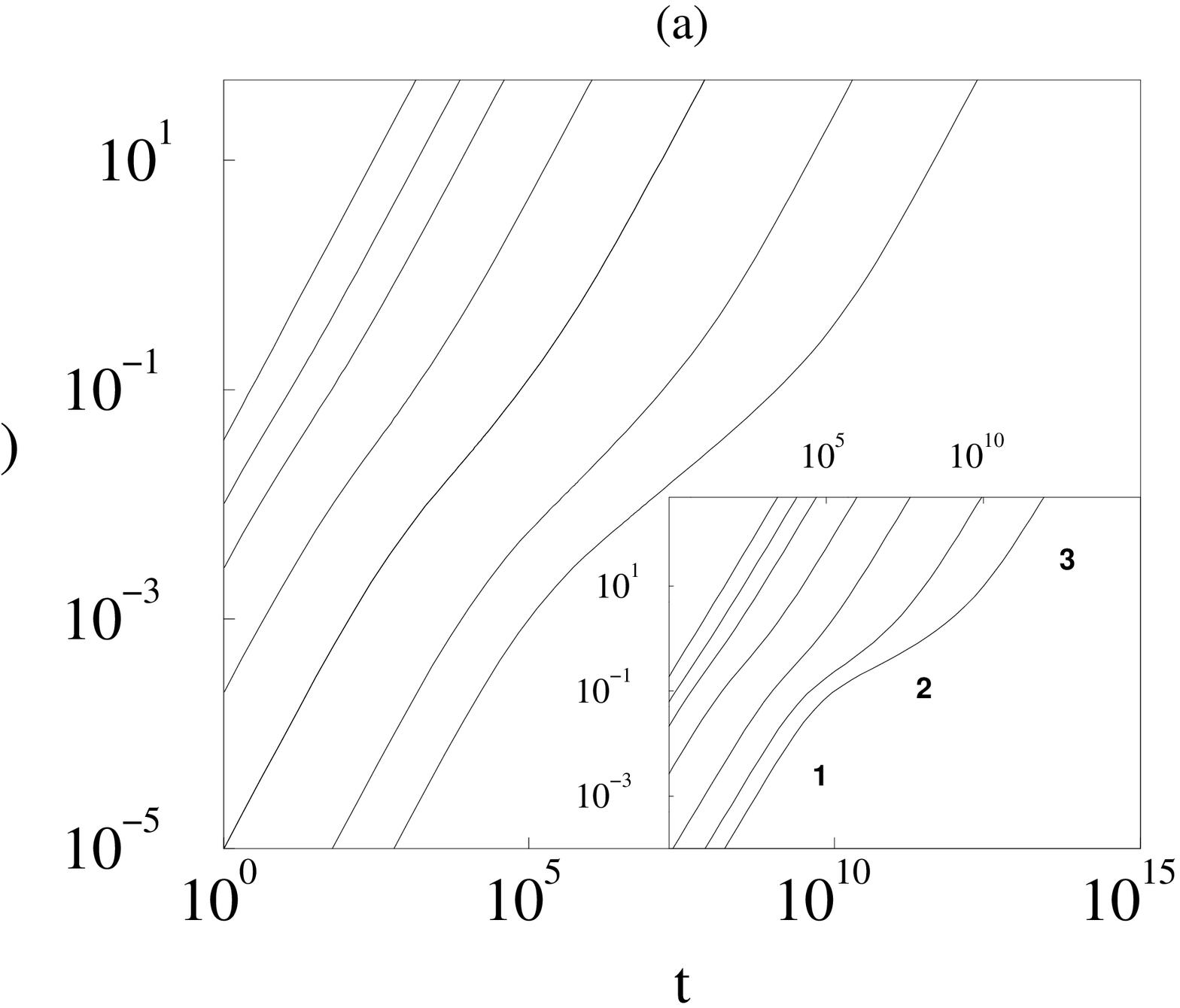}}
\centerline{\epsfxsize=.9\hsize \epsffile{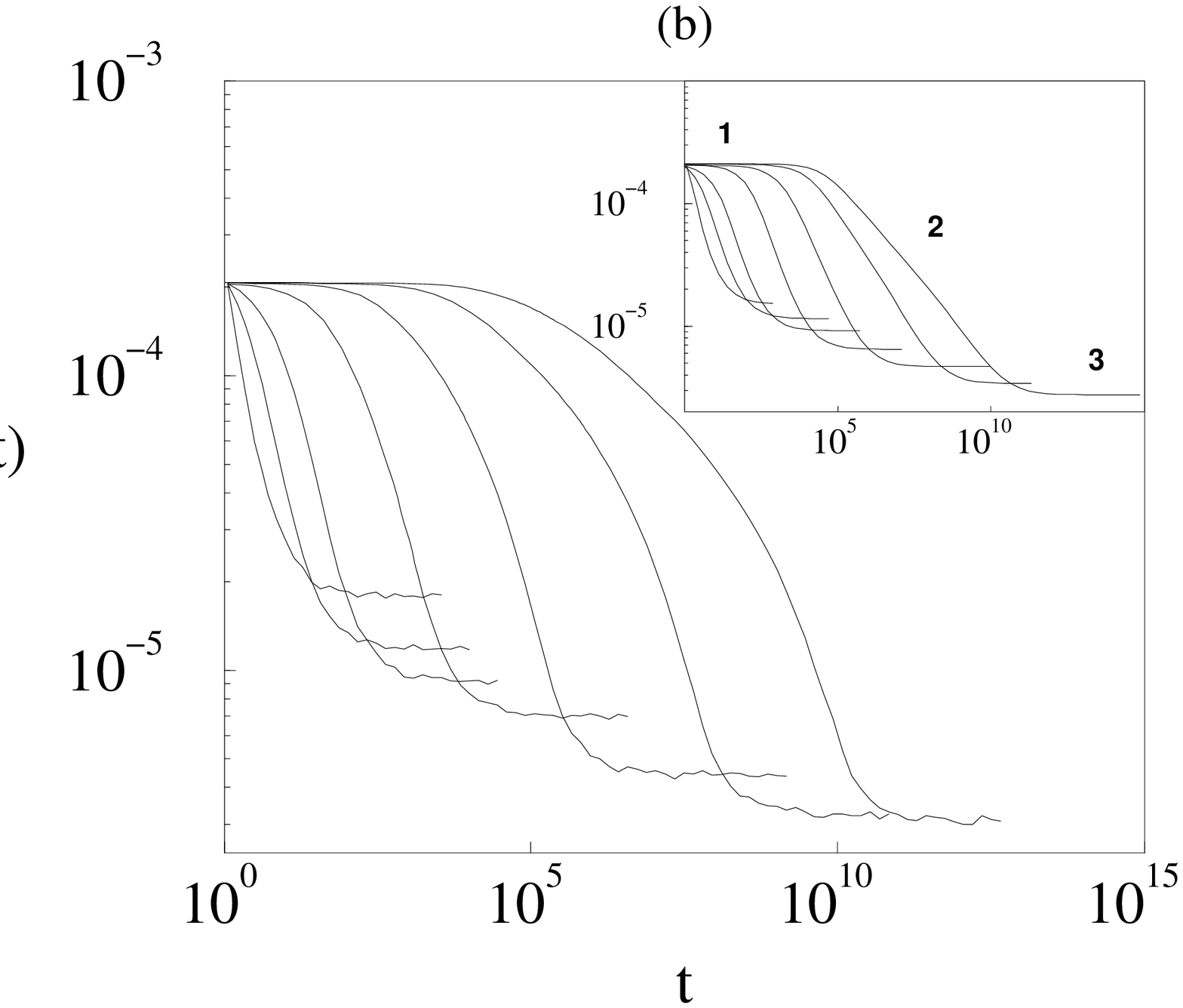}}
\caption{$R^2(t)$ (a) and $S^2(t)$ (b) 
are plotted for a quench to $T=10^{-4}$
and densities $\overline \rho =$ 0.70, 0.80, 0.85, 0.90, 0.93, 0.95,
0.96 (from left to right). 
The main figures show the outcome of the numerical 
simulation, the insets refer to the mean field
approximation. The location of the three regimes described in the text 
are outlined by the numbers 1, 2 and 3.}
\label{1}
\end{figure}

\begin{figure}
\centerline{\epsfxsize=.9\hsize \epsffile{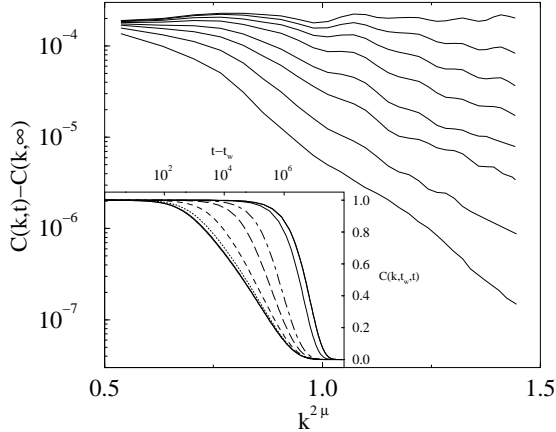}}
\caption{The slow decay, as a function of $k$, 
of $C(\vec k,t)$ in the second regime, is 
plotted here against $k^{2\mu}$, with $\mu=1/6$, for increasing times.
$C(\vec k,t)$ is consistent with a non Gaussian fit:
$C(\vec k,t)\simeq {\cal C} e^{-(l(t)k)^{2\mu}}$.
The inset shows the decay of $\tilde C(\vec k,t',t)$ as a function of 
$t-t'$ in mean field
for a quench to $T=10^{-4}$ with $\overline \rho =0.95$. Different
$t'$ are shown ($t' = 1, 10^4, 10^5, 10^6, 10^7, 10^9, 10^{11}, 
10^{13}$, from left to right; the last two curves collapse 
because the stationary state is entered and time translational invariance 
is obeyed).}
\label{3}
\end{figure}

The decay of the average density fluctuations 
$S^2(t)=\sqrt{<(\rho-\overline \rho )^2>}$, is plotted in Fig.~$1^b$.
For low densities $\overline \rho$ a {\it normal liquid} region is observed;
$S^2(t)$ very quickly decays to the constant value characteristic of
the equilibrium state.
By raising the density, one enters a different region.
Here the vanishing of $M$ slows the dynamics: this
produces a long-lasting off-equilibrium glassy behavior before
the equilibrium state is reached.
The behavior of $S^2(t)$ shows that for large densities the 
dynamics can be divided in three regimes. 
Initially, for $t$ smaller than a characteristic time $\tau _p$, 
$S^2(t)$ remains constant. This is the first 
regime.
An analysis of the system configuration $\rho (\vec r,t)$ in this time
domain does not show any appreciable evolution: 
the system is pinned in the initial configuration.
Then, for $t> \tau _p$ 
a second regime is entered and less dense regions equilibrates whereas high 
density zones are still practically frozen. This is characterized
by the decrease of $S^2(t)$.
In this regime, that will be referred to as {\it slow evolution},
one observes pronounced correlated spatial heterogeneities in the
system. 
This spatial pattern is outlined by a slow decay, as a function of $k$, of the 
structure factor $C(\vec k,t)=<\rho (\vec k,t)\rho (-\vec k,t)>$,
that is consistent with a stretched exponential fit (see Fig.~\ref{3})
$C(\vec k,t)\simeq {\cal C} \exp {-[l(t)k]^{2\mu}}$
with $\mu \simeq 1/6$ (at variance with the Gaussian decay of standard 
diffusion), similarly to some experimental observations \cite{mallamax}.
Finally the system enters the equilibrium state characterized by 
a constant value of $S^2(t)$.
This whole pattern is reflected by the behavior of 
the particle mean square displacement $R^2(t)$, shown in Fig.$1^a$, calculated
through
\begin{equation}
R^2(t)=\int _0 ^t D(t')dt' 
\label{R(t)}
\end{equation}
where $D (t)=<M(\rho)>$ is the average mobility \cite{new_nota}.
For low densities $R^2(t)\sim t$ in the whole time domain, as expected
for simple diffusion. As the density is increased toward the limiting value
$\overline \rho =1$ three regimes can again be distinguished. After an initial
linear increase (regime 1) a progressively more pronounced 
inflection is observed
in an intermediate time domain (regime 2) whose duration is 
enhanced as $\overline \rho$ is increased. 
The same pattern is observed  
in both spin-glass like 
lattice gas and Lennard-Jones molecular dynamics simulation \cite{kob}.
Asymptotically, in equilibrium, $R^2(t)\sim t$, as for simple diffusion.

In order to get analytical results we now introduce an approximation 
on Eq.~(\ref{diff}) by first expanding the logarithm on the r.h.s. of 
Eq.~(\ref{diff}) to lowest order \cite{notaD} and then by replacing 
the mobility $M (\rho )$ with the effective diffusivity
$D (\rho)=<M(\rho)>$. 
Since average
quantities do not depend on the position, due to space homogeneity, 
one has $D(\rho)=D(t)$. Eq.~(\ref{diff}) then becomes
\begin{equation}
\frac {\partial \rho (\vec r,t)}{\partial t}=D(t)\nabla ^2 
\rho (\vec r,t)+\eta (\vec r,t)
\label{meanf}
\end{equation}
where the rescaling  
$t\to t /
[\overline \rho (1 - \overline \rho)]$ and 
$T\to {\tilde T} =T \overline \rho (1 - \overline \rho)$,
has been performed, and
$<\eta (\vec r,t)\eta(\vec r',t')>=-2{\tilde T}D(t)\nabla ^2
\left [ \delta (\vec r-\vec r')\delta (t-t')  
\right ] $. 
Transforming Eq.~(\ref{meanf}) into reciprocal space, 
one obtains the following formal solution
for the two time correlator 
$C(\vec k,t',t)=
<\rho (\vec k,t')\rho (-\vec k,t)>$, ($t\geq t'$) 
\begin{eqnarray}
C(\vec k,t',t) &=& e^{-[R^2(t')+R^2(t)]k^2} \nonumber \\
&& \left \{C(\vec k,0,0)+\tilde T\left [ e^{2R^2(t')k^2}-1 \right ]\right \}
\label{formal}
\end{eqnarray}
The whole problem is now reduced to the knowledge of $R(t)$ which 
must be calculated self-consistently enforcing Eq.~(\ref{R(t)}),  
where $D(t)$ is given by
$D(t)=<M(\rho)>=\int _0 ^{1} M(\rho)P(\rho)d\rho$.
Here $P(\rho)$ is the probability distribution of the density field that, 
for Eq.~(\ref{meanf}) can be shown~\cite{gauss} to be Gaussian at all times.
Then we have
\begin{equation}
D(t)=[2\pi S^2(t)]^{-1/2}\int _0 ^{1} 
M(\rho)e^{-(\rho -\overline \rho)^2/[2S^2(t)]}d\rho
\label{effdiff}
\end{equation}
The quantity $S(t)$ can be computed as  
$S^2(t)=(2\pi )^{-d}\int _{k<\Lambda} C(\vec k,t,t)d\vec k $ where
$\Lambda$ is a momentum cutoff of order $a^{-1}$.
From Eq.~(\ref{formal}), $S(t)$ is a function of $R(t)$:
\begin{equation}
S^2(t)=
h[S^2(0) - q {\tilde T}]
R^{-d}(t)\Phi _d [\sqrt {2}\Lambda R(t)]+q{\tilde T}
\label{esse}
\end{equation}
where 
$\Phi _d[x]=\int _0 ^x y^{d-1}\exp (-y^2) dy$, 
$q=(\Sigma _d /d)[\Lambda /(2\pi)]^d$, $h=[d/(\Lambda \sqrt 2)^d]$ and
$\Sigma _d $ is the surface of the $d$-dimensional unitary hypersphere. 
Notice that the asymptotic value of the density fluctuations
$S^2(\infty )=qT\overline \rho (1-\overline \rho)$ vanishes
at the point of structural arrest.

Eqs.~(\ref{R(t)}, \ref{effdiff}, \ref{esse}) are a closed set of equations 
that can be studied analytically. 
In the present approximation the non linearity of $M(\rho)$ is accounted 
for by a self-consistency prescription for the 
calculation of $D(t)$. Similar approximation techniques are well developed
and widely used in several field of statistical physics \cite{appro}
producing reliable results \cite{notaq}.
From Eq.~(\ref{formal}) the normalized correlator
$\tilde C(\vec k,t',t)=C(\vec k,t',t)/C(\vec k,t',t')$ is given by
\begin{equation}
\tilde C(\vec k,t',t) = e^{-[R^2(t)-R^2(t')]k^2}
\label{cnorm}
\end{equation}
showing that a scaling form
$\tilde C(\vec k,t',t)={\cal S}[\phi (t)/\phi(t')]$, with $\phi (t)=
\exp \{-R^2(t)k^2\}$, is obeyed as suggested by
a scaling approach to dynamical processes \cite{NC_sca}. 

An important issue to understand the off-equilibrium dynamics is the 
relation between the response function to a small 
perturbing field $h_{\vec k}$, 
$\chi_{\vec k}(t',t)\equiv \int _{t'} ^t d\tau 
\frac{-\delta <\rho_{\vec k}(t)>}
{\delta h_{\vec k}(\tau)}$, 
and the correlation function in the unperturbed situation, 
$C(\vec k,t',t)$. In equilibrium systems, where the fluctuation-dissipation 
theorem holds, the quantity
$X\equiv {\tilde T} \partial \chi_{\vec k}(t',t)/\partial C(\vec k,t',t)$ 
is equal to one. Out of equilibrium this relation is 
violated, and, generally, $X$ is a function of $t'$ and $t$: 
$X=X(t',t)$ \cite{BCKM}.
In the present mean-field approximation, one may interestingly show that the  
generalized ``fluctuation-dissipation" ratio (FDR) $X$ is a function 
of the sole $t'$: 
$X(t')=\{[C(\vec k,0,0)-\tilde T]\exp{(-2k^2R^2(t'))}+\tilde T\}^{-1}$.
Only if $t'\rightarrow\infty$, the usual version of the FDR  
with $X=1$ is recovered (notice that $X\leq 1$). 

In the following we will report the main results of the mean field analysis 
referring to a longer publication~\cite{lungo} for all the details.
From the solution of the model one sees that if the density $\overline \rho$
is small or the temperature ${\tilde T}$ is high one immediately enters the 
asymptotic stationary state that will be described later on. 
For high densities and low temperatures, on the other hand, the evolution 
remains markedly far from equilibrium for a long period and three dynamical 
regimes are found corresponding to different behaviors of $R(t)$ 
(see Figs.~$1$), as discussed below.

{\bf Regime 1 - Pinning: }
For short times, such that $R(t)\ll \Lambda 
^{-1}$, as shown by Eq.~(\ref{cnorm}), 
$\tilde C(\vec k,t',t)$ is essentially 
constant since  $|k| < \Lambda$ and the system looks pinned. In this 
regime we also have 
$\Phi _d[\sqrt {2}\Lambda R(t)]\sim R(t)^d$, consequently
$S(t)\simeq S(0)$ and $D(t)\simeq D(0)$. Therefore  
$R^2(t)\simeq D(0)t$ (see inset Fig.~$1^a$). 
The duration of this regime is  $\tau _p \simeq \Lambda ^{-2} D^{-1}(0)$.
Physically $\tau _p $ corresponds to the time the particle
spends inside its cell (cage).

{\bf Regime 2 - Slow evolution:} Pinning lasts up to $\tau_p$. 
For $t >\tau_p$, we have $R(t)> \Lambda $, thus particles diffuse out of 
the cages and the evolution starts. For sufficiently long times, computing 
$R(t)$ through Eqs.~(\ref{R(t)}, \ref{effdiff}, \ref{esse}) 
one finds $R^2(t)\simeq b(\ln t)^\delta$ (see inset Fig.~$1^a$), 
where $b=const.$  
and $\delta=6/d$. Eq.~(\ref{formal}) implies that, for 
fixed $t'$ the correlator decays as an enhanced power law 
(see inset Fig.~\ref{3})
\begin{equation}
\tilde C(\vec k,t',t)= \exp \{R^2(t')k^2\}
\exp \left \{ -b [\ln (t)]^\delta k^2 \right \}
\label{enhanced}
\end{equation}
and $S(t)\sim (\ln t)^{-3/2}$ (see inset Fig.~$1^b$).
A logarithmic relaxation of the density fluctuations 
is also observed in Molecular Dynamics simulations of out of 
equilibrium liquid glass formers \cite{Rieger}.
When also $t'>\tau_p$, one has 
$\tilde C(\vec k,t',t)= 
\exp \left \{ -b ([\ln (t)]^\delta-[\ln (t')]^\delta) k^2 \right \}$. 
The characteristic duration time, $\tau_e$, of the 
slow evolution regime can be estimated \cite{lungo}, at low ${\tilde T}$, to be 
$\tau_e\sim M^{-1}(\overline \rho)$.

{\bf Regime 3 - Stationary state:} For long times  $t > \tau_e$ a simple
diffusive behavior is obtained because $D(t)$ always attains asymptotically 
a constant value $D(\infty)$. This imply $R^2(t)\simeq D(\infty)t$, as can 
be seen in Fig.~$1^a$, so that the normalized correlator exhibits the usual 
exponential decay as a function of $t$: 
$\tilde C(\vec k,t',t)=e^{-D(\infty)tk^2}e^{\{R^2(t')k^2\}}$ 
(see inset Fig.~\ref{3}). 
When $t'>\tau_e$, we have 
$\tilde C(\vec k,t',t)=e^{-D(\infty)[t-t']k^2}$
and time translational invariance is obeyed. In the small temperature limit 
the density fluctuations $S(\infty)$ can be approximatively neglected 
\cite{notadded} and  
$D(\infty)\simeq M(\overline \rho)$. This leads to 
$\tau _0 = M^{-1}(\overline \rho)$ which gives Eq.~(\ref{tau0}), as
previously stated.  

So far we have studied the out off-equilibrium
evolution of a system governed by Eq.~(\ref{diff}) in the presence of
a vanishing mobility for which Eq.~(\ref{tau0}) holds in equilibrium.
We also want to consider the case  in which the mobility vanishes as 
a power law, as is found for instance in the Mode-Coupling Theory of 
supercooled liquids \cite{gotze}: $M(\rho)=[1-\rho (\vec r,t)]^\gamma$.
This relation implies an algebraic divergence of $\tau_0$: 
$\tau _0 =M(\overline \rho )^{-1}$, as experimentally found in supercooled
liquids in the temperatures or densities regions far away the 
ideal glassy transition. 
The different form of the mobility, as stated before, does not changes
the global picture described so far with three different regimes. 
In the first and third regimes $R^2(t)$ is linear in $t$ as before, 
while in the second one we find, for $\gamma >1$, an anomalous diffusion 
$R(t)\simeq wt^\beta$, with $w$ = const. and $\beta=4/(\gamma d+4)$. 
In this regime we also find \cite{lungo} a stretched exponential decay 
of the normalized correlator
$\tilde C(k,t',t)\sim \exp \left \{
R^2(t') k^2 \right \}
\exp \left \{
- w t^\beta k^2 \right \}$.

In this paper we have introduced a phenomenological equation 
for off equilibrium glassy dynamics.
The only ingredients of the model are the diffusive behavior and the
request of a Vogel-Fulcher (or algebraic) divergence of
$\tau_0 $, obtained by assuming a mobility as in Eq.(\ref{mob2}).
With these sole ingredients the out of equilibrium evolution of the model
is observed to be highly non trivial, even in the mean field approximation 
which we have studied in details. 
Consistently with mean field theory, also the numerical integration of the 
full model shows the existence of a gradual crossover from a 
{\it normal liquid } to a glassy behavior 
by raising the density. Some properties that are
observed in systems close to the glassy transition, 
such as the existence  of strong spatial heterogeneities, 
anomalous diffusion, slow decay and aging of 
density autocorrelations, are exhibited by the model.
These predictions, as long as the non trivial fluctuation dissipation 
ratio $X(t')$, are all amenable of experimental check.
In mean field this whole richness is observed in the preasymptotic 
off equilibrium dynamics (which, however, may be exponentially long), 
whereas the asymptotic  
equilibrium evolution is trivial. This is an important difference with 
real glassy systems, where a non trivial decay of $\tilde C
(\vec k, t',t)$ is also observed in equilibrium.
However in mean field a non exponential decay of $\tilde C(\vec k, t',t)$
can be ruled out on general grounds due to the Doob's theorem \cite{markov}.
Further studies are in progress in order to 
characterize the complicate fluctuations occurring
in the equilibrium state.

{\bf Acknowledgments} 
We are grateful to M.Zannetti for interesting discussions
and to S.Roux for valuable comments on the manuscript. 
F. C. thanks M. Cirillo and R. Del Sole for their hospitality 
in Rome university.
This work was supported with the TMR network contract
ERBFMRXCT980183 and by MURST(PRIN 97).


\begin{references}
\bibitem{notaD} Due to our approximation on $F[\rho]$, $\rho $ 
might become larger than $1$. Thus we take $M(\rho)$ zero outside 
the interval $[0,1]$.
\bibitem{notaq}

This scheme produces a poor approximation of the spatial 
configuration of the system, as in testified by the gaussian
$k$-dependence of $C(\vec k,t)$; the time-dependence, however, 
is well reproduced, as can be tested by a comparison with the numerical 
simulations presented in Figs.~(\ref{1},\ref{3}).
\bibitem{notadded}

Since we are interested in the region close to the point of structural arrest, 
where, as already noticed, $S^2$ vanishes, this approximation is justified.
\end{references}
\end{document}